# 3rd Workshop on Hybrid Development Approaches in Software System Development

Paolo Tell[1(✉)], Stephen MacDonell[2], and Sherlock A. Licorish[3]

[1]*IT University of Copenhagen, Copenhagen, Denmark*
pate@itu.dk
[2]*Auckland University of Technology, Auckland, New Zealand*
stephen.macdonell@aut.ac.nz
[3]*University of Otago, Dunedin, New Zealand*
sherlock.licorish@otago.ac.nz

**Abstract**

*Evidence shows that software development methods, frameworks, and even practices are seldom applied in companies by following the book. Combinations of different methodologies into home-grown processes are being constantly uncovered. Nonetheless, an academic understanding and investigation of this phenomenon is very limited. In 2016, the HELENA initiative was launched to research hybrid development approaches in software system development. This paper introduces the 3rd HELENA workshop and provides a detailed description of the instrument used and the available data sets.*

## 1. INTRODUCTION

A software process is the game plan to organize project teams and run projects. Even though a multitude of development methods and frameworks have been proposed over the years, the daunting statement that "there is no silver bullet" [3] serving all possible setups still holds strong. Given the context of a company, project, or team, the selection of the appropriate development approach or the creation of an ad-hoc combination is still a challenge. Recent research as well as experience from practice shows companies utilizing different development approaches to assemble the best-fitting approach for the respective company.

After West identified in 2011 [10] the presence of what he labelled water-scrum-fall—the ad-hoc combination of different software process philosophies into home-grown instances to fit the different organizational needs of a company—several researchers investigated this phenomenon. In 2015, a systematic review to reveal the current state of practice in software process use revealed a considerable imbalance between the research understanding of practice and practice itself [9]. Consequently, the HELENA initiative was born.

In the remainder of this paper we describe in more details the HELENA project (Sect.2) and the third instance of the yearly HELENA workshop (Sect.3). Section4 concludes this paper by providing a summary of future activities.

## 2. THE HELENA STUDY

HELENA is an international exploratory multistage survey-based study on the use of "**H**ybrid d**E**ve**L**opm**EN**t **A**pproaches in software systems development". In [7], we defined *hybrid software development approaches* as any combination of agile and traditional (plan-driven or rich) approaches that an organizational unit adopts and customizes to its own context needs (e.g., application domain, culture, processes, project, organizational structure, techniques, technologies, and other factors).

**The Team and its Organization**
After three years, the HELENA project now involves about 80 researchers from (currently) 25 countries. Besides the member role, the structure of the project comprises a core group tasked with ensuring the progression of activities, and a main representative for each of the 25 countries responsible for maximising local coordination. The project aims to investigate the current state of practice in software and systems development; in particular: which development approaches (traditional, agile, main-stream, or home-grown) are used in practice and how they are combined, how such combinations were developed over time, as well as if and how standards (e.g., safety standards) affect the development process as such and the methods applied. With this information, we aim to push forward systematic process design and improvement activities to allow for more efficient and reduced-overhead development approaches.

**The Data Collection Instrument**
To achieve these goals the project is designed to collect data through a survey, which has been refined over several iterations. After being successfully tested within Europe in project stage one [7], the HELENA project is reaching the end of stage two, during which the survey has been conducted globally in more than 25 countries. A third and final stage will conclude the project. In stage three, focus groups will perform in depth research on community-defined topics of interest based on the results of stage two. The form and coverage of the questionnaire that has been used in stage two can be seen in Tables 1 and 2. These tables not only show the question and answer types, but also the arrangement of questions into questionnaire pages, which is an important piece of information when it comes to understanding the different data sets that have been created for data analysis. These are discussed in the remainder of this section together with a selection of descriptive results.

Table 1. Detailed overview of the HELENA instrument structure. ([n]: number of available options; FT: free text; SC: single choice, MC: multiple choice; RT: rating; Llex: Likert scale including "don't know" option.
Additional information on the options provided for each question including details on the Likert scales and rating variables are publicly available at https://goo.gl/yoA1m4)

| Page | Questions | Code | Type |
|---|---|---|---|
| 1 | *Introduction* | I001 | |
| 2..14 | *Main questionnaire (see Table 2)* | | |
| 15 | • Do you have any further comments or issues not addressed so far? | C001 | FT |
| | • If you want to be informed about the study's outcomes and possible future iterations (with in-depth interviews), please leave your e-mail address here: | C002 | FT |
| | • In future iterations, we plan to complement this survey with in-depth interviews. Would you be willing to participate in these interviews? | C003 | SC[2] |
| | • Have you already participated in stage 1 of the HELENA survey? | C004 | SC[2] |
| | • Have you filled in the questionnaire more than once (i.e., for more than one project/product)? | C005 | SC[2] |
| | • How did you learn about this survey/how were you contacted? | C008 | SC[5]+FT |
| 16 | • For which company/organization do you work? | C006 | FT |
| | • Are we allowed to name your company in the list of participants? | C007 | SC[3]+FT |
| 17 | *Closing* | I004 | |

**The Data Sets**
The survey instrument was accepting responses between May and November 2017. The survey was promoted through personal contacts of the 75 participating researchers, through posters at conferences, and by posts to mailing lists, social media channels (Twitter, Xing, LinkedIn), professional networks and websites (ResearchGate and researchers (institution) home pages).

In total, 1467 data points were collected, of which 691 are complete. As a first step, given the discrepancy between the two sets, the data was analyzed by two members of the core team to investigate the level of completeness, which yielded the identification of a third set that was deemed complete enough to pursue the majority of the investigations planned. The constraints applied during this process were mainly based around the presence of core questions (see Table 2):

PU09: used frameworks and methods,
PU10: used practices,
PU04: self-awareness regarding the use of hybrid approaches, and
PU05: self-assessment of philosophies followed with respect to the general project related activities listed in the Guide to the SWEBoK [1].

Following this rationale and to avoid bias, rather then applying filters to the data set on the presence of answers—as participants were given the option to skip questions—we identified page #9 of the questionnaire as the marker that had to be reached for a data point to be accepted to the third data set. This set comprises the data points that are considered usable for the majority of the investigations that the HELENA project planned to research, for the main objective is eventually to explore hybrid development approaches in software system development. Therefore, as visually represented in Fig.1, members of the HELENA team were given access to three data sets[1]:

**Full:** comprising 1467 data points. Members were discouraged from using this set.

**Suggested:** comprising 732 data points[2] selected according to the process described above.

**Completed:** comprising 691 data points of participants that answered the questionnaire in its entirety.

**Selected Results from the Suggested Dataset**
The remainder of this section presents some selected results from the *Suggested dataset* to showcase the richness of the information collected through the survey.

D001 - **Company size** (n = 732) Five categories were provided to choose from: *micro* (<10 employees) (11.6%),

---
[1] Only the core team has access to the survey instrument given that page #16 contained confidential information that the instrument tool collected separately.

[2] Five data points were additionally dropped due to the instrument marking such entries as inconsistent.

Table 2. Detailed overview of the HELENA instrument questions and variables. ([n]: number of available options; FT: free text; SC: single choice, MC: multiple choice; RT: rating; LIex: Likert scale including "don't know" option. Additional information on the options provided for each question including details on the Likert scales and rating variables are publicly available at https://goo.gl/yoA1m4)

| Page | Questions | Code | Type |
|---|---|---|---|
| 2 | • What is your companys size in equivalent full-time employees (FTEs)? | D001 | SC[5] |
|  | • What is the main business area of your company? | D002 | MC[7]+FT |
| 3 | • Please describe the project or product your answer is related to in a few words (less than 100), or provide an acronym. | D008 | FT |
|  | • What is the size of the project or product to which your answer is related? | D009 | SC[5] |
|  | • Is the project or product your answer refers to carried out in a (globally) distributed manner? | D003 | SC[4] |
|  | • In which country are you personally located? | D004 | SC[241] |
|  | • What is the major role you have in this project or product? | D007 | SC[11]+FT |
|  | • How many years of experience do you have in software and systems development? | D010 | SC[5] |
| 4 | • What is the target application domain of the project or product your answer is related? | D005 | MC[19]+FT |
|  | • In the project or product you refer to, a software failure conceivably can: <criticality> | D006 | MC[9]+FT |
| 5 | • Does your company define a company-wide standard process for software and system development? | PU01 | SC[3] |
|  | • How was your project-specific development approach defined? | PU08 | SC[6]+FT |
|  | • Do you intentionally deviate from defined policies? | PU11 | SC[2]+FT |
| 6 | •Which of the following frameworks and methods do you use? | PU09 | RT[24] |
|  | • Do you use further frameworks and methods? | PU14 | FT |
| 7 | • Which of the following practices do you use? | PU10 | RT[36] |
|  | • Do you use further practices? | PU15 | FT |
| 8 | • Do you combine different development approaches in the development of one project or product? | PU04 | SC[2] |
| 9 | • For the following standard activities in the project or product development, please indicate to which degree you carry out these activities in a more traditional or more agile manner. | PU05 | LIex[11] |
| 10 | • How were the combinations of development frameworks, methods, and practices in your company developed? | PU07 | MC[3]+FT |
|  | • What are the overall goals that you aim to address with your selection and combination of development approaches? | PU12 | MC[18] |
|  | • Is there a further/other motivation to combine the different development frameworks, methods, and practices? | PU06 | FT |
| 11 | • To what degree did the combination of approaches help you to achieve your goals? | PU13 | RT[18] |
| 12 | • Do you implement external standards in your company? | PS01 | SC[2]+FT |
| 13 | • Why have you implemented the aforementioned standards? | PS02 | MC[3]+FT |
|  | • How is the compliance of the development process assessed? | PS03 | MC[5]+FT |
|  | • Does agility challenge the implementation of the standards you have to apply? | PS04 | SC[2]+FT |
|  | • Is the project or product your answer relates to also subject to certification? | PS05 | SC[2]+FT |
| 14 | • Based on your personal experience, please rate the following statements: | EX01 | LIex[8] |
|  | • Based on your personal experience, please specify any problems that have arisen regarding your current process and your current application domain. | EX02 | FT |

*small* (11–50 employees) (13.0%), *medium* (51–250 employees) (24.9%), *large* (251–2499 employees) (27.0%), *very large* (>2500 employees) (23.1%). Among the respondents, 0.4% did not answer this question. An interesting aspect about this distribution lies in the fact that, after merging the micro and small categories, the groups become extremely balanced in size.

**D009 - Product/Project size** (n = 732) Again five categories were provided: *very small* (<2 person weeks) (2.2%), *Small* (2 person weeks - 2 person months) (3.8%), *medium* (2 person months - 6 person months) (15.0%), *large* (6 person months - 1 person year) (18.7%), *very large* (>1 person year) (60.2%). All respondents answered.

**D010 - Experience** (n = 732) Among the respondents, the majority reported more than 10years of experience (59.8%). Following, 18.3% reported 6–10 years experience, 14.1% between 3 and 5years, 5.1% 1–2 years, and only 2.7% stated less then one year.

**D003 - Distribution** (n = 732) Given the significant implications, an interesting aspect of software system development regards whether teams are physically co-located. Respondents were asked to describe the level of distribution of their product/project. Also for this variable, the distribution appears to provide good sample sizes. In particular, of the products/projects: 37.6% are *co-located*, 24.6% *are distributed nationally (within the same country)*, 11.9% are *distributed regionally (within the same continent)*, and 25.8% are *globally distributed*. One respondent (0.1%) did not answer.

## 3. THE WORKSHOP

Continuing along the tradition of yearly meetings and the community work initiated at ICSSP 2016 (Austin, Texas), ICSSP 2017 (1st workshop, Paris, France [8]), and Profes 2017 (2nd workshop, Innsbruck, Austria [6]), the 3rd HELENA workshop focuses on discussing results from Stage 2 of the HELENA project.

In this workshop, we aim to bring together all (academic) contributors and further interested people to: (i) report the current state and (tentative) outcomes of the HELENA survey (from a global and regional perspective); (ii) develop a work program and define next steps within the whole community; and, (iii) build working groups to research on selected (sub-)topics of interest.

**Workshop Organization**
The 3rd HELENA workshop is a 1-day workshop aimed at bringing together all members of the HELENA project to network and work together around selected topics. An overview of the schedule is provided in Table 3. We will start by providing an overview of the current state of the project including the main objectives that have been achieved and chief results that have been found. Ample space will then be given to the presentation of submitted articles and reports from ongoing activities. In this regard, two pieces of work have been accepted to the workshop:

1. Using Institutional Theory as a lens, the author models the tension between traditional software engineering and agile software development of today's software engineering, allowing a better understanding of how and why hybridisation comes about in software organisations. The paper provides an ideal ground for an engaged discussion at the workshop. [2]

2. A second paper reports on the potential relationship between institutional goals and the adoption of certain software development methods in German organisations. Ultimately, it finds that no such relationships exist at a broad level—the most often cited goals are common irrespective of methods used. The paper succeeds in presenting open possibilities for strengthening the analysis, and the investigation relating criticality of the product with the level of agility presented in the paper represents indeed an interesting discussion topic for the workshop. [4]

The second segment of the event will focus on the presentation of major topics that have been identified through the project and that will lead to further endeavours. Active sessions will be run to critically scrutinize such topics and, if relevant, identify new ones. The segment will close with a plenary session to consolidate the outcomes. Finally, the third segment will revolve around the discussion and refinement of the next steps of the HELENA project—chiefly Stage three.

## 4. CONCLUSION AND FUTURE WORK

Over a small timeframe, the work conducted by the HELENA community has managed to provide significant evidence highlighting the importance of this topic. Several pieces of work have already been published in highly relevant venues (e.g., Profes [9], ICSSP [7], IEEE Software [5]). Insofar, we have shown that hybrid development approaches in software system development are a reality that affect companies regardless of size and industry sector. We have also, in several instances, characterized the evidence based on different regions.

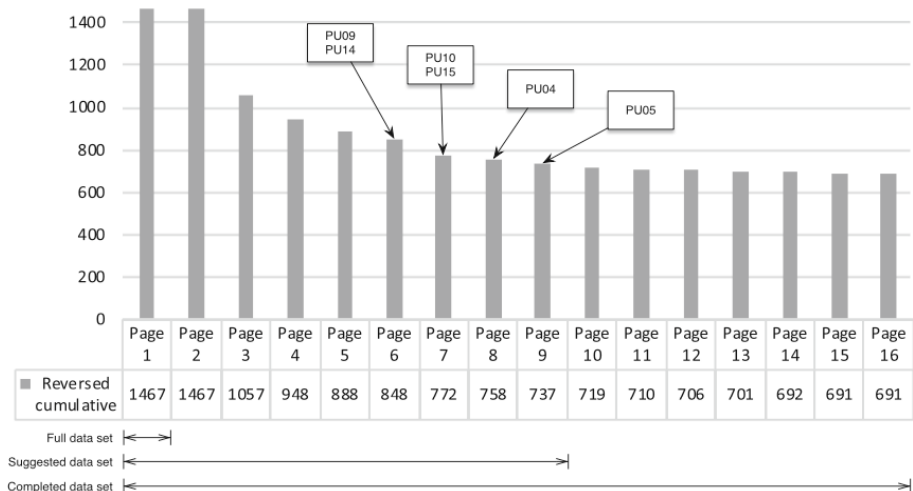

Fig. 1. Overview of response rate per page detailing the size of the three data sets.

Current activities within the HELENA community are investigating, inter alia, (i) the impact of the strategies used to devise hybrid development approaches on the ability to achieve set goals, and (ii) the alignment of software and system development frameworks, methods, and practices taught in higher education with those used in industry. Through more thorough and rigorous analysis of the large collections of data, we are now uncovering an increasing number of results that not only constantly strengthen past results, but allow us to push forward systematic process design and improvement activities geared towards more efficient and reduced-overhead development approaches.

**Acknowledgments.** We want to thank Profes 2018 organization board for providing us with the opportunity to hold the third HELENA workshop in conjunction with Profes 2018. We look forward to continuing such fruitful collaboration with the Profes community.

# REFERENCES

[1] Bourque, P., Fairley, R.E.: Guide to the Software Engineering Body of Knowledge (SWEBOK(R)): Version 3.0. IEEE Computer Society Press, Washington, D.C. (2014)

[2] Doležel, M.: Possibilities of applying institutional theory in the study of hybrid software development concepts and practices. In: Kuhrmann, M., et al. (eds.) PROFES 2018. LNCS, vol. 11271, pp. 441–448. Springer, Cham (2018)

[3] Fraser, S., Mancl, D.: No silver bullet: software engineering reloaded. IEEE Softw. 25(1), 91–94 (2008). https://doi.org/10.1109/MS.2008.14

[4] Klünder, J., et al.: Towards Understanding the Motivation of German Organizations to Apply Certain Software Development Methods. In: Kuhrmann, M., et al. (eds.) PROFES 2018. LNCS, vol. 11271, pp. 449–456. Springer, Cham (2018)

[5] Kuhrmann, M., et al.: Hybrid software development approaches in practice: a European perspective. IEEE Softw., 1 (2018). https://doi.org/10.1109/MS.2018. 110161245

[6] Kuhrmann, M., Diebold, P., MacDonell, S., Münch, J.: 2nd workshop on hybrid development approaches in software systems development. In: Felderer, M., Méndez Fernández, D., Turhan, B., Kalinowski, M., Sarro, F., Winkler, D. (eds.) PROFES 2017. LNCS, vol. 10611, pp. 397–403. Springer, Cham (2017). https://doi.org/10. 1007/978-3-319-69926-4 28

[7] Kuhrmann, M., et al.: Hybrid software and system development in practice: waterfall, scrum, and beyond. In: Proceedings of the 2017 International Conference on Software and System Process, ICSSP 2017, pp. 30–39. ACM, New York (2017). https://doi.org/10.1145/3084100.3084104, http://doi.acm.org/ 10.1145/3084100.3084104

[8] Kuhrmann, M., Münch, J., Tell, P., Diebold, P.: Summary of the 1st international workshop on hybrid development approaches in software systems development. ACM SIGSOFT Softw. Eng. Notes 42(4), 18–20 (2018)

[9] Theocharis, G., Kuhrmann, M., Münch, J., Diebold, P.: Is Water-Scrum-Fall reality? On the use of agile and traditional development practices. In: Abrahamsson, P., Corral, L., Oivo, M., Russo, B. (eds.) PROFES 2015. LNCS, vol. 9459, pp. 149–166. Springer, Cham (2015). https://doi.org/10.1007/978-3-319-26844-6 11

[10] West, D., Gilpin, M., Grant, T., Anderson, A.: Water-scrum-fall is the reality of agile for most organizations today. Forrester Research, Cambridge (2011)